\documentclass[a4paper,english,aps,prb,twocolumn]{revtex4}
\usepackage[T1]{fontenc}
\usepackage[latin9]{inputenc}
\usepackage{textcomp}
\usepackage{bashful}
\usepackage{bm}
\renewcommand{\bm}{\mathbf}
\usepackage{relsize}
\usepackage{amsmath}
\usepackage{graphicx}
\usepackage{bm}
\usepackage{color}
\newcommand{\LP}{{\mathrm{L1}}_0}

\newcommand{\MS}{{M_{\mathrm s}}}
\newcommand{\TC}{{T_{\mathrm c}}}
\usepackage{dcolumn}
\usepackage{bm}
\usepackage{doi}
\usepackage{hyperref}

\graphicspath{{figures/}}

\makeatother

\usepackage{babel}

\makeatother

\usepackage{babel}
\begin{document}
\title[]{Temperature Dependent Ferromagnetic Resonance via the Landau-Lifshitz-Bloch Equation: Application to FePt}

\author{T.~A.~Ostler}
\affiliation{$^1$ Department of Physics, University of York, York YO105DD, U.~K.}
\author{M.~O.~A.~Ellis}
\affiliation{$^1$ Department of Physics, University of York, York YO105DD, U.~K.}
\author{D.~Hinzke}
\affiliation{$^2$ Fachbereich Physik, Universit\"{a}t Konstanz, D-78457 Konstanz, Germany}
\author{U.~Nowak}
\affiliation{$^2$ Fachbereich Physik, Universit\"{a}t Konstanz, D-78457 Konstanz, Germany}

\begin{abstract}
Using the Landau-Lifshitz-Bloch (LLB) equation for ferromagnetic materials, we derive analytic expressions for temperature dependent absorption spectra as probed by ferromagnetic resonance (FMR). By analysing the resulting expressions, we can predict the variation of the resonance frequency and damping with temperature and coupling to the thermal bath. We base our calculations on the technologically relevant $\LP$ FePt, parameterised from atomistic spin dynamics simulations, with the Hamiltonian mapped from \emph{ab-initio} parameters. By constructing a multi-macrospin model based on the LLB equation and exploiting GPU acceleration we extend the study to investigate the effects on the damping and resonance frequency in $\mu$m sized structures. 
\end{abstract}
\maketitle

\section{Introduction} 
The magnetic properties of ferromagnetic structures such as thin films, nanowires and nanoparticles
have been studied extensively both experimentally~\cite{Clinton2008,Schellekens2013} and theoretically~\cite{Krone2011,Sukhov2008}. The interest in
these particles is driven by fundamental features on the one hand and
technological perspectives on the other~\cite{Sun2000,Cowburn1999,Dormann2007BOOK}. Ferromagnetic resonance (FMR), which
has been applied with great success to thin ferromagnetic films in the
past~\cite{Farle1998}, can be used to measure important material properties, such as the damping, gyromagnetic ratio and anisotropy constant. The temperature dependence of these properties for large or complex structures are often difficult to predict using analytical treatments, especially when temperature effects are included~\cite{Sukhov2008,Ellis2012}. As well as being difficult to calculate analytically, temperature dependent calculations of (for example) FMR can be slow to converge. The convergence can become particularly troublesome if thermal fluctuations are accounted for. A specific motivation for this work is the interest in $\LP$ FePt materials, which is a promising candidate for ultrahigh density magnetic recording~\cite{Weller1999,Rottmayer2006}.

The ability to tune magnetic properties, such as the damping is important for example, in devices based on spin transfer torque where a low damping of a free layer is essential for reducing the power consumption and can affect the signal to noise ratio~\cite{Smith2001}. In some cases such as in GMR read sensors, high damping is preferred to improve thermal stability~\cite{Maat2008}.

For technologies based on Heat Assisted Magnetic Recording (HAMR), understanding temperature effects and fluctuations in strongly anisotropic materials, will be crucially important. In this article we present analytical and numerical calculations of the material properties of strongly anisotropic materials at elevated temperatures. We do so by utilizing the formalism of the Landau-Lifshitz-Bloch (LLB) equation of motion for ferromagnetic particles, which has an intrinsic temperature dependence via various input functions. In the first part of the paper we present the derivation of the temperature dependent analytic expression for the power absorbed by the particle. This analytic expression allows us to look at the effect of temperature on FMR curves for single domain particles. The temperature dependent input functions that enter into the LLB formalism have been parameterised from  atomistic spin dynamics with the exchange parameters calculated directly from \emph{ab-initio} calculations~\cite{Kazantseva2008}. We have tested the expressions with a single spin and multispin (with exchange) LLB numerical model, by showing a number of resonance curves at different temperatures against the derived expressions (without demagnetizing fields). The analytic expressions for the damping and resonance frequency show the overall trend of the temperature dependent behaviour.

In the second half of the paper we extend the scope of our analysis using a multi-macrospin model based on the LLB formalism with large number of exchange coupled macrospins. We present numerical calculations of FMR in 2D and 3D structures with the inclusion of demagnetizing effects and (stochastic) thermal fluctuations. Specifically, we have investigated the effects of the anisotropy constant and film thickness and anisotropy on the measured damping in out-of-plane films at high temperatures. Our findings show that, depending on thickness or anisotropy, there is a competition between the demagnetizing and anisotropy energy that can modify the damping significantly. We have implemented this large scale model on the CUDA GPU platform so that even with the inclusion of the stochastic thermal terms, it is possible to obtain good averaging of the FMR power spectra.

There are limited experimental ferromagnetic studies of chemically ordered FePt due to it's large magnetocrystalline anisotropy~\cite{Alvarez2013}. However, it is possible to perform so-called optical FMR with the use of laser pulses~\cite{Becker2014}. In a theoretical work by Butera~\cite{Butera2006} the resonance spectra were calculated using a computational model for disordered nanoparticles of FePt. This study showed that the measured damping depended strongly on the amount of disorder. To our knowledge there are no systematic studies on the temperature dependence of the properties such as damping due to the limited fields in typical FMR setups, our results provide insight into this complex issue.

\section{Landau Lifshitz Bloch Equation}
\label{s:llb} 

The LLB equation for magnetic macrospins describes the time evolution of an ensemble of atomic spins and allows for relaxation of the magnitude of the magnetization. The equation was originally derived by Garanin~\cite{Garanin1997} within a Mean-Field approximation from the
classical Fokker-Planck equation for atomistic spins interacting
with a heat bath.  The resulting LLB equation
has been shown to be able to describe linear domain walls, a domain
wall type with non-constant magnetisation length. These results are
consistent with measurements of the domain wall mobility in YIG
crystals close to the Curie point ($\TC$)~\cite{Kotzler1993} and with
atomistic simulations~\cite{Kazantseva2005}. Furthermore, the
predictions for the longitudinal and transverse relaxation times
have been successfully compared with atomistic simulations~\cite{Chubykalo-Fesenko2006}. Consequently, we use this equation in the
following for the thermodynamic simulations of macro-spins. The use of the LLB formalism has the advantage over traditional micromagnetics that it automatically allows for changes in the modulus of the magnetisation. In theory it is indeed possible to calculate temperature dependent FMR using the atomistic spin dynamics (ASD) model however, such an approach would be extremely computationally expensive. This computational expense in the ASD model arises because, for FMR calculations, large system sizes are required to reduce the effects of thermal noise. Whilst large systems are possible to calculate, the FMR calculations also require averaging over many cycles of the driving field, up to 100's of nanoseconds. These two restrictions combined means that this method is not suitable, even with GPU acceleration or a (for example MPI) distributed memory solution~\cite{Evans2014}.

A further challenge for accurate calculation of magnetic properties is the accounting of the long-ranged exchange in materials such as FePt. Through proper parameterisation of the LLB equation~\cite{Kazantseva2008} one can account for such long ranged interactions in the \emph{so-called} multiscale approach~\cite{Kazantseva2008}. Via this multiscale approach we can then bridge the gap between electronic structure calculations to large scale (of the order of micrometres) calculations of material properties. With this in mind the LLB model is then ideally placed to describe temperature dependent ferromagnetic resonance.

The LLB equation, without the stochastic term, can be written in the form:
\begin{eqnarray}
&&\bm{\dot{m}}=-\gamma \lbrack \bm{m}\times  \bm{H}_{%
\mathrm{eff}} ]+\frac{\gamma \alpha _{||}}{m^{2}}\left(%
\bm{m}\cdot  \bm{H}_{\mathrm{eff}}\right) %
\bm{m} \nonumber \\
&&\qquad {}-\frac{\gamma \alpha
_{\perp}}{m^{2}} \left[\bm{m}\times \left \lbrack
\bm{m}\times \bm{H}_{\mathrm{eff}} %
\right] \right].
\label{e:llb}
\end{eqnarray}
Besides the usual precession and relaxation terms, the LLB equation contains
another term which controls longitudinal relaxation (second term in equation~\ref{e:llb}). Hence ${\bf m}$ is a spin
polarisation which is not assumed to be of constant length and even its
equilibrium value, $m_{\rm e}(T)$, is temperature dependent. The value of $\bf m$ is equal to the ratio of the magnetization of the macrospin normalised by the magnetization at saturation ($\bm{M}/M_{\mathrm{s}}$). ${\alpha_{\parallel}}$ and
$\alpha_{\perp}$ are dimensionless longitudinal and transverse damping
parameters (defined below) and $\gamma$ is the gyromagnetic ratio taken to be the free electron value. The transverse damping parameter in this equation is related to what is usually measured in experiments (the Gilbert damping, $\alpha_{\mathrm{g}}$) by the expression:
\begin{equation}
\alpha_{\mathrm{g}}=\frac{\alpha_{\perp}}{m}
\label{eq:dampGLLB}
\end{equation}
The LLB equation is valid for finite temperatures and even above $\TC$, though the damping parameters and effective fields are different in the two regions. Throughout this paper, we are only interested in the case $T \leq \TC$ with the damping parameters $ \alpha_{\parallel} = \frac{2\lambda  T}{3 \TC}$ and $ \alpha_{\perp} = \lambda \Big(1 - \frac{T}{3 \TC}\Big)$. The single particle free energy (without demagnetizing fields) is given by:
\begin{eqnarray}
f &=& -B\MS^0m_z+\frac{\MS^0}{2\tilde{\chi}_{\perp}}(m_x^2+m_y^2) \nonumber \\
&&+ \frac{\MS^0}{8\tilde{\chi}_{\|}m_e^2}(m^2-m_e^2)^2,
\end{eqnarray}
and the effective fields, $ \bm{H}_{\mathrm{eff}} = - \frac{1}{M_{\mathrm s}^0} \frac{\delta f}{\delta \bf{m}}$  given by \cite{Garanin1997}:
\begin{equation}
  \bm{H}_{\mathrm{eff}} = \bm{B}+\bm{H}_{A}+
       \frac{1}{2\tilde{\chi}_{\Vert }}\left(1-\frac{m^{2}}{m_{\rm e}^{2}}\right) \bm{m},
    \label{e:Heffm}
\end{equation}
where $\bm{B}$ represents an external magnetic field and 
 $\bm{H}_{A}  =  -\left( m_{x}\bm{e}_{x}+m_{y}\bm{e}_{y}\right)/\tilde{\chi}_{\perp}$  an anisotropy field. Here, the susceptibilities $\tilde{\chi}_{l}$ are defined by 
$\tilde{\chi}_{l} = \partial m_l / \partial H_l$, where $H_l$ is the $l={\parallel,\perp}$. In these equations, $\lambda$ is a microscopic parameter which
characterizes the coupling of the individual, atomistic spins with
the heat bath. For the purpose of testing the model we use a thermal bath coupling constant of $\lambda$=0.05, consistent with Ref.~\onlinecite{Alvarez2013}. There are differing values of the damping constant in the literature, for example, for granular FePt Becker \emph{et al} measured a damping constant of 0.1 using an optical FMR technique, whereas Alvarez \emph{et al} found a value of 0.055 using standard FMR in a broad frequency range~\cite{Alvarez2013}. It should be pointed out here that whilst $\lambda$ is a coupling to the thermal bath equivalent to that used in atomistic spin dynamics and is assumed to be temperature independent.

At this point we should take some time to define the different constants related to the damping and their differences. The parameters, $\lambda$, $\alpha_{\perp}$, $\alpha_{\parallel}$ and $\alpha_{\mathrm{g}}$ correspond to the thermal bath coupling, the temperature dependent transverse and longitudinal damping parameters and the damping parameter that one would measure experimentally, respectively. The thermal bath coupling is temperature independent and is a phenomenological parameter that is the same as that used in atomistic spin dynamics. The transverse and longitudinal damping parameters that enter into the LLB equation define the relaxation rates of the transverse and longitudinal magnetization components. Finally, the parameters, $\alpha_{\mathrm{g}}$, is equal to the perpendicular damping ($\alpha_{\perp}$) that enters into the equation of motion, divided by the magnetization and is what one would find in an FMR measurement from the linewidth.

For the application of this equation one has to know \emph{a-priori} the spontaneous equilibrium magnetisation $m_{\rm e}(T)$, the perpendicular ($\tilde{\chi}_{\perp}(T)$) and parallel ($\tilde{\chi}_{\parallel}(T)$) susceptibilities. In this work these are calculated  separately from a Langevin dynamics simulation of an atomistic
spin model, however it is possible to calculate these properties from mean field calculations~\cite{Mendil2013}. We use a model for FePt which was introduced earlier and which is meanwhile well-established in the literature \cite{Mryasov2005,Nowak2005,Hinzke2007,Hinzke2008}. Since this model was derived from first principles, a direct link is made from spin dependent density functional theory calculations, via a spin model, to our macrospin simulations. The calculation of these parameters is discussed in more detail in Ref.\onlinecite{Kazantseva2008}.

\section{Analytic Solution for the FMR Absorbed Power Spectrum P($\omega$)}
\label{sec:pw}

The focus of this section is on the derivation of an analytical solution for the power spectrum $P(\omega)$ using the LLB equation for a single macro spin.  The power, $P(\omega)$, absorbed in an FMR experiment is given by~\cite{Sukhov2008}:
\begin{equation}
P(\omega) = \langle \mathbf{M} \cdot \frac{\partial \mathbf{B}}{\partial t} \rangle=-\frac{\omega}{2 \pi}  \int_0^{2\pi/\omega}M_{\mathrm{S}}V m_x \dot{B_x} {\rm d} t,
\label{e:defP}
\end{equation}
where $V$ is the volume of the macrospin and $\omega$ is the frequency of the driving field. The right hand side of equation~\ref{e:defP} assumes that the time-varying field is applied in the $x$ direction with the static applied field in $z$. The time-dependence of the x-component of the magnetization can be derived from Eq.~\ref{e:llb}. Using the assumptions that 
$m^2$ is constant and $m_x$ as well as $m_y$ are small, leading leads to the approximation $m_z \approx m$.  Under this assumption Eq.~\ref{e:llb} can be written in 
linearised form. Together with the linearised form of the effective field the solution of the resonance frequency, $\omega_0$, and transverse relaxation time, $\tau$, can be obtained (for full details see appendix~\ref{sec:app2}):

\begin{eqnarray}
 \omega_0(T)  &=& \gamma \Big(B_z  + \frac{m(T)}{\tilde{\chi}_{\perp}(T)}\Big) ,
 \label{e:res1}
 \end{eqnarray}
 \begin{eqnarray}
\tau(T) & =& \frac{m(T)}{\lambda \big( (1 - \frac{T}{3 \TC})\omega_0(T) - \frac{2}{3} \gamma \frac{T}{\TC} {H}_{\mathrm{eff}}(T)^z \big)}.
 \label{e:res2}
\end{eqnarray}
Here $m=m_e + \tilde{\chi}_{\parallel}B_z$ is an approximation written to first order of the susceptibility for the purposes of the analytic calculation.
In the zero temperature case under the conditions that $\alpha = \alpha_{\perp}$, $\alpha_{\parallel} = 0$ and $m = m_{\rm e} =1$, $\omega_0$ and $\tau$ are the same as for the Landau Lifshitz Gilbert (LLG) equation, 
$\omega_0  = \gamma (B_z  +  \frac{1}{\tilde{\chi}_{\perp}})$  and  
$\tau   =  \frac{1}{\lambda \omega_0}$.

\begin{figure}[htbc!]
\begin{center}
\includegraphics[width=\columnwidth]{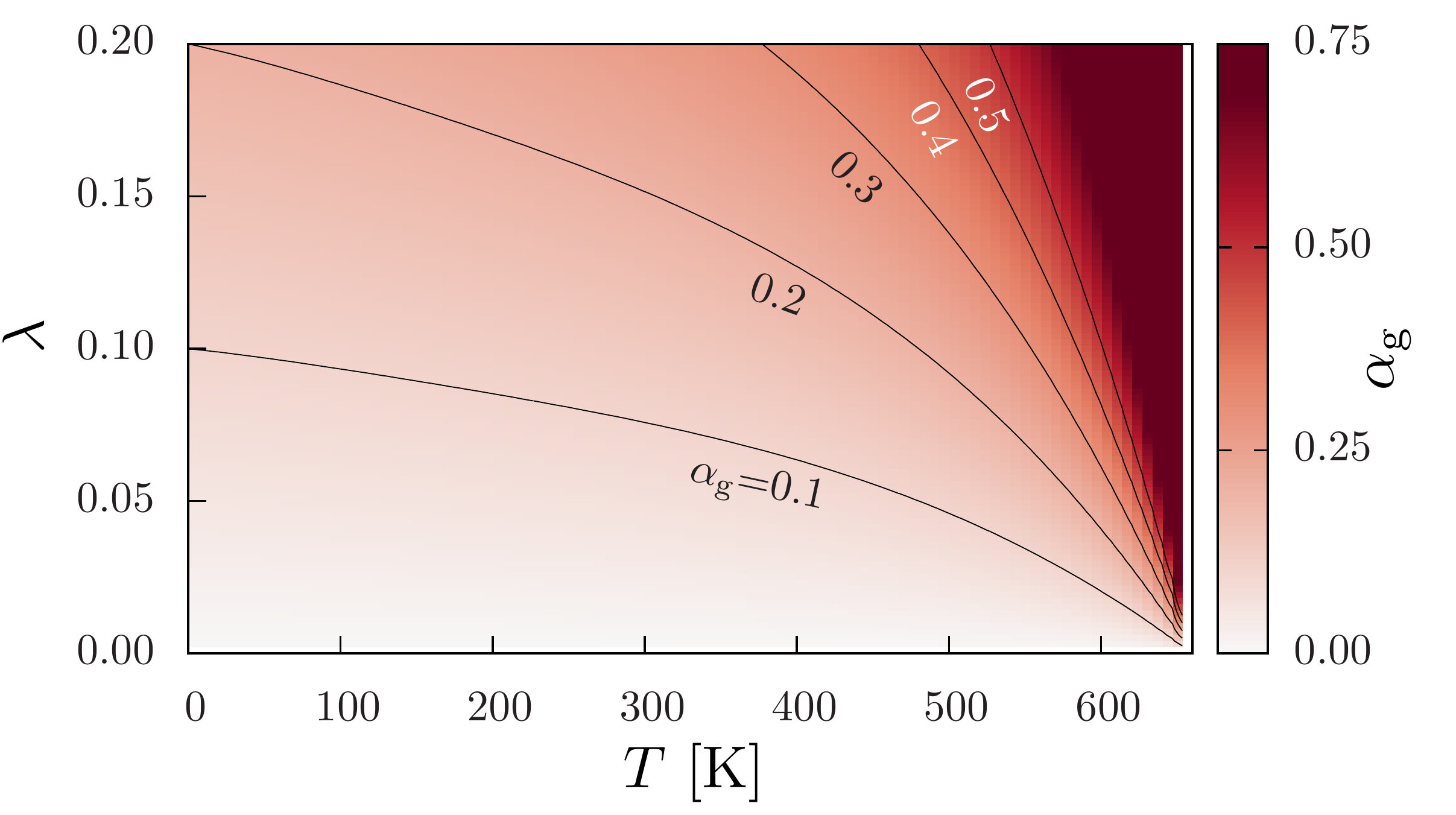}
\caption{(Color online) Analytically derived Gilbert damping as a function of temperature and the intrinsic coupling to the thermal bath, $\lambda$, valid for a single macrospin without demagnetizing effects. For each value of $\lambda$ the damping is shown to increase with temperature consistent with other works~\cite{BeckerAPL2014,Chubykalo-Fesenko2006}. The lines are contours of constant measured damping.}
\label{fig:GDA}
\end{center}
\end{figure}

The analysis of equations~\ref{e:res1} and~\ref{e:res2} show that there is little variation of the measured damping, $\alpha_{\mathrm{g}}$, with the applied field as one would expect~\cite{BeckerAPL2014}. Also, at low temperature, as expected, the measured damping equal to the input coupling to the thermal bath, $\lambda$. The temperature dependence of $\alpha_{\mathrm{g}}$ shows that (for a chosen value of $\lambda$) there is an increase with temperature, diverging at the Curie point. In a recent article~\cite{BeckerAPL2014} the measured damping as a function of applied field (up to 7~T) was shown to be almost independent of temperature. In the same study, the damping was measured at two temperatures; 170~K and 290~K. Between these two temperatures the damping was shown to be around 0.1 with a slight increase as one would expect. Figure~\ref{fig:GDA} shows the analysis of equation~\ref{e:res1} and~\ref{e:res2} for the physical input parameters for FePt. In the figure the measured damping is calculated as $\alpha_{g}=1/\omega_0 \tau$ and is shown to increase with temperature, diverging at the Curie point. The contours show lines of constant damping explicitly. This demonstrates that if one assumes no temperature dependence of the thermal bath coupling, $\lambda$, the measured damping will not be constant.

The solution of the resulting inhomogeneous differential equation~\ref{e:DGLm1}, combined with equation~\ref{e:defP} leads us to the analytic solution for the power absorbed during ferromagnetic resonance as a function of the frequency of the driving field:
\begin{eqnarray}
P(\omega,T)  &=& \frac{\mu_{\rm s} \omega^2}{4}  \frac{\gamma \alpha_{\perp} B_0^2}{\frac{1}{\tau^2} + (\omega - \omega_0)^2},
\label{e:LLBP_red}
\end{eqnarray}
where the temperature dependence comes from $\omega_0$ and $\tau$ (see equations~\ref{e:res1} and~\ref{e:res2}) and $B_0$ is the amplitude of the driving field. In the zero temperature case this solution reduces to that from the LLG equation.

The analytic solution given by equation~\ref{e:LLBP_red} can be compared to the numerical results, by integration of the LLB equation and using equation~\ref{e:defP}. By applying an alternating driving field in the $x$ direction and averaging equation~\ref{e:defP} until convergence we can compare the results of a single spin to the analytic expression. For FePt there is a strong uniaxial exchange anisotropy, therefore in the absence of any static applied field we still see a very strong FMR line for single domain particles. Throughout the calculations we use a driving field amplitude ($B_0$) of 0.005T and a static applied field ($B_z$) of 1~T. We integrate the LLB equation using the Heun numerical scheme with a 5~fs timestep.

\begin{figure}[htbc!]
\begin{center}
\includegraphics[angle=0,width=\columnwidth]{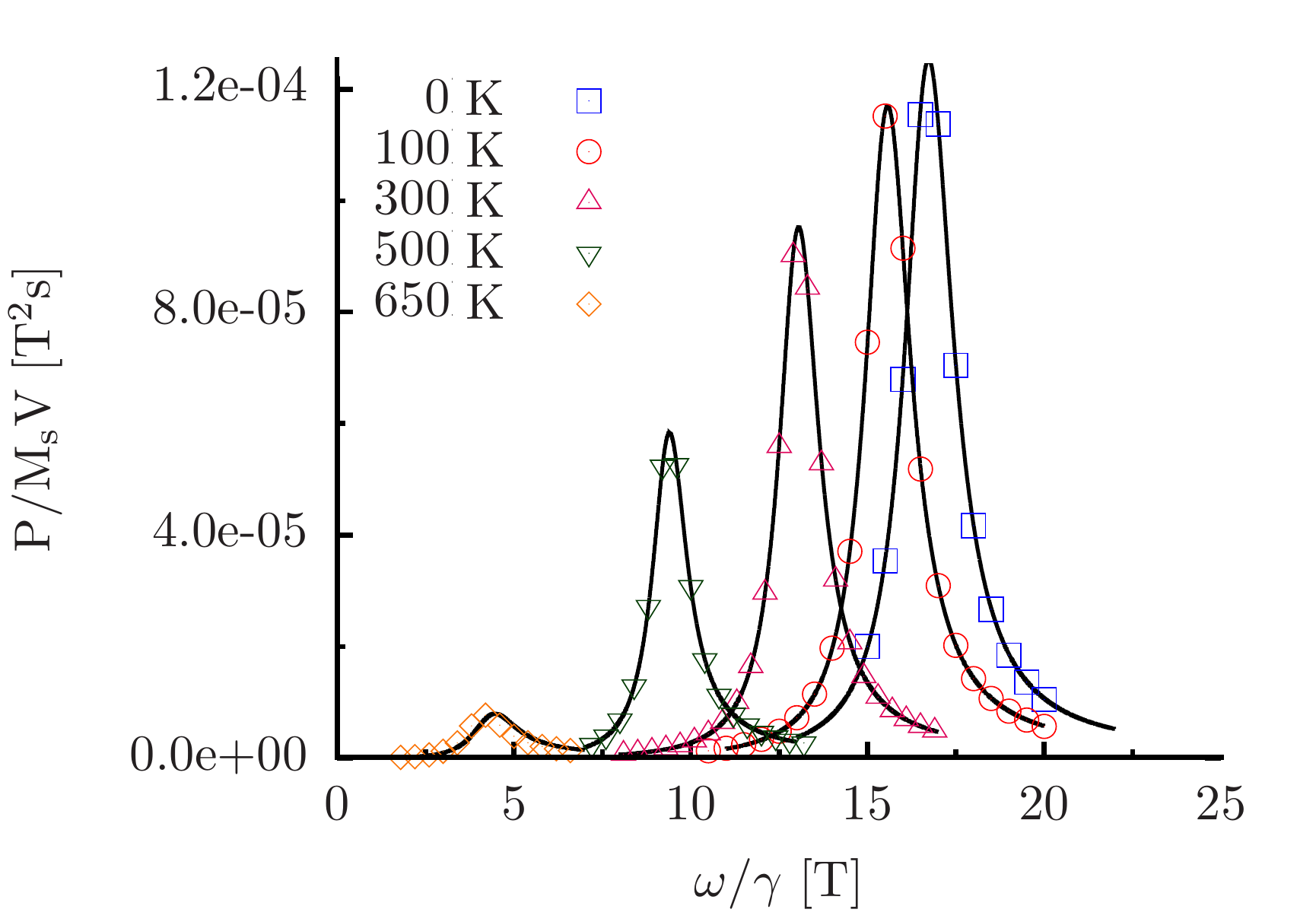}
\caption{(Color online) Power spectrum vs frequency in a 1~T applied field. The data points are from LLB simulations for a single macrospin and the solid lines are given by Eq.~\ref{e:LLBP_red}.}
\label{f:Bz_0T}
\end{center}
\end{figure}

Figure~\ref{f:Bz_0T} shows the calculated absorbed power as a function of frequency for a range of temperatures using the single spin LLB model.
As we can see from Figure~\ref{f:Bz_0T} there is a large decrease in the resonance frequency, given by equation~\ref{e:res1}, which we would expect to occur because of the decrease in the anisotropy field. The analytic solution agrees perfectly with the numeric model, except as we approach the Curie temperature. This is because in the analytical treatment we approximate the magnetisation in a field, $B_z$, to depend on the parallel susceptibility ($m=m_e+\tilde{\chi}_{\parallel} B_z$) which diverges as we approach the Curie temperature. This point has been discussed in appendix~\ref{sec:app2} and is an error in the analytic treatment only, not in the form of the LLB equation. 
\begin{equation}
P(\omega) = P_0 \frac{\omega^2}{(\omega \tilde{\alpha}_g)^2 + (\omega-\tilde{\omega}_0)^2}
\label{eq:fit}
\end{equation}
where $\tilde{\alpha}_g$, $P_0$ and $\tilde{\omega}_0$ are free fitting parameters and we use the tilde to distinguish the resonance frequency and damping from the analytically derived values. This use of this fitting procedure is allows us to compare with experimental observations as this would be the kind of analysis required to extract the damping parameter ($\alpha_{\mathrm{g}}$).

\section{Multi-Macrospin Numerical Results}
\label{sec:mmm}

In the following section we introduce the stochastic LLB equation that introduces thermal fluctuations. As well as the normal terms in the LLB described by equation~\ref{e:Heffm}, we also include exchange coupling between the macrospins and the magnetostatic fields. The LLB equation with stochastic thermal terms included is written for each spin, $i$, in the form:
\begin{eqnarray}
&&\bm{\dot{m}}_i=-\gamma \lbrack \bm{m}_i\times  \bm{H}^i_{%
\mathrm{eff}} ] + \boldsymbol{\zeta}_{i,\parallel} \nonumber \\
&&\qquad {}-\frac{\gamma \alpha
_{\perp}}{m_i^{2}} \left[\bm{m}_i\times \left \lbrack
\bm{m}_i \times (\bm{H}^i_{\mathrm{eff}}+\boldsymbol{\zeta}_{i,\perp}) %
\right] \right] \nonumber \\
&&\qquad {}+ \frac{\gamma \alpha _{||}}{m_i^{2}}\left(%
\bm{m}_i\cdot  \bm{H}^i_{\mathrm{eff}}\right) %
\bm{m}_i.
\label{e:sllb}
\end{eqnarray}
The stochastic fields, $\boldsymbol{\zeta}_{i,\perp}$ and $\boldsymbol{\zeta}_{i,\parallel}$ have zero mean and the variance~\cite{Evans2012}:
\begin{eqnarray}
\langle \zeta_{i,\perp}^{\eta}(0)\zeta^{\theta}_{j,\perp}(t)\rangle &=& \frac{2|\gamma|k_{\mathrm{B}}T(\alpha_{\perp}-\alpha_{\parallel})}{M_s V \alpha_{\perp}^2}\delta_{ij} \delta_{\eta \theta}\delta(t) \nonumber \\
\langle \zeta_{i,\parallel}^{\eta}(0)\zeta^{\theta}_{j,\parallel}(t)\rangle &=& \frac{2|\gamma|k_{\mathrm{B}}T\alpha_{\parallel}}{M_s V}\delta_{ij} \delta_{\eta \theta}\delta(t)
\label{eq:meanvar}
\end{eqnarray}
where $\parallel$ is the additive noise, $\eta$ and $\theta$ represents the Cartesian components. As well as the stochastic field, the exchange is also included in the form:
\begin{equation}
\bm{H}_{{ex}}^i = \frac{A(T)}{m_e^2} \frac{2}{M_s^0 \Delta^2} \displaystyle\sum_{j \in neigh(i)} (\bm{m}_j - \bm{m}_i)
\label{eq:LLBExch}
\end{equation}
where $A(T)$ is the exchange stiffness, $\Delta$ is the cell length and $M_s^0$ is the saturation magnetization.

It should be pointed out that the inclusion of the stochastic term into the LLB equation leads to a slightly reduced $T_{\mathrm{C}}$ as compared to the LLB without the stochastic term~\cite{Evans2012}.

\begin{figure}[htbc!]
\begin{center}
\includegraphics[angle=0,width=\columnwidth]{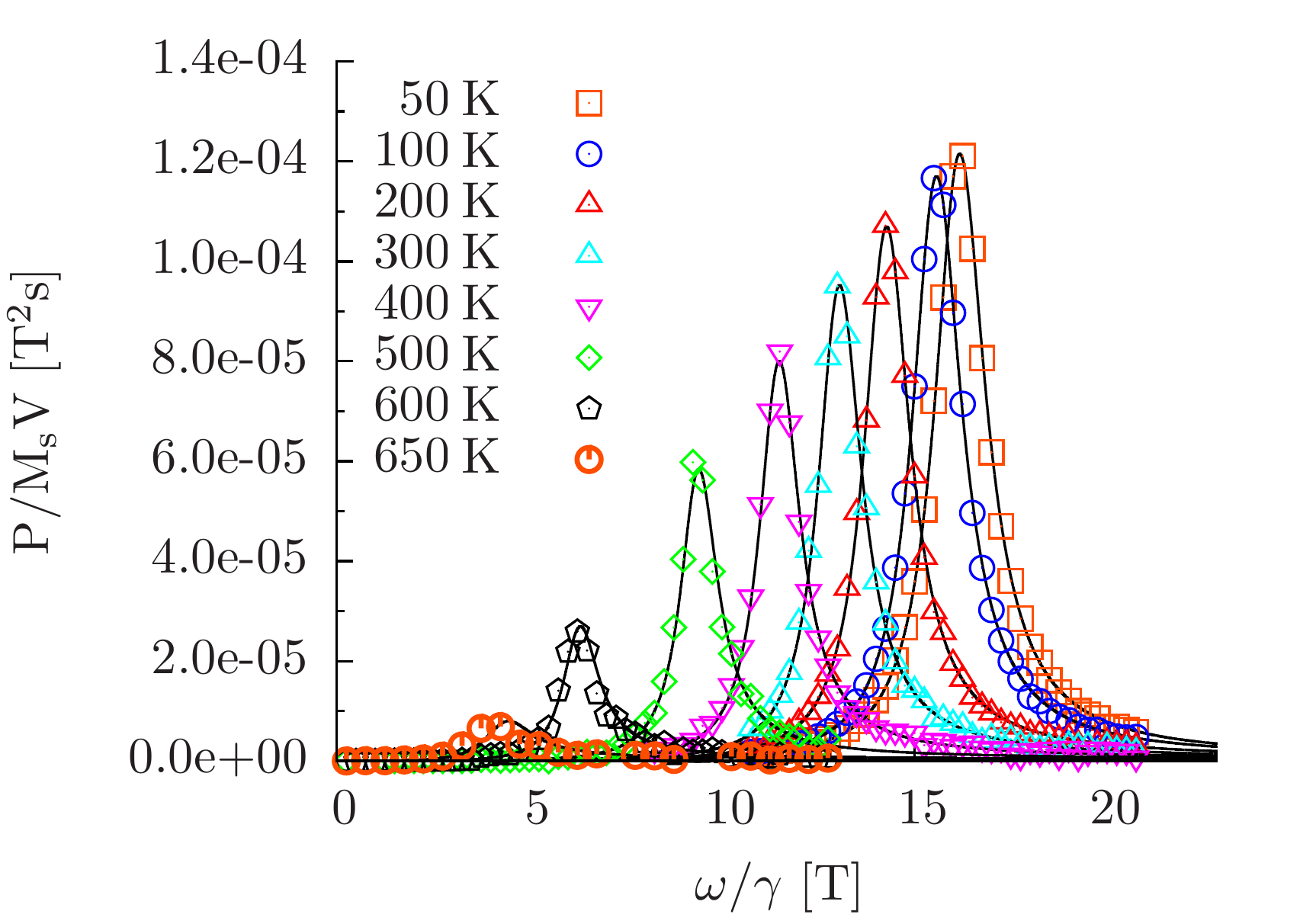}
\caption{(Color online) Power spectrum vs frequency in a 1T applied field. The data points are from LLB simulations for many exchange coupled macrospins including the stochastic fields and exchange and the solid lines are given by Eq.~\ref{e:LLBP_red} (no magnetostatic fields are included here).}
\label{fig:multiFMR}
\end{center}
\end{figure}

Figure~\ref{fig:multiFMR} shows the power spectrum as a function of frequency for multi-macrospin calculations (coupled by exchange) for a system size of (100~nm)$^3$ with a unit cell discretization of (6.25~nm)$^3$, though we have checked unit cell sizes down to (3.125~nm)$^3$, i.e. below the typical domain wall size of 4-6~nm. The solid lines are the analytical solution, equation~\ref{e:LLBP_red}.

As discussed in the introduction, we have also introduced into our model demagnetizing effects to extend the analytic study to more realistic materials. We have taken the approach of that of Lopez-Diaz \emph{et al.} used in the GPMagnet software~\cite{Lopez-Diaz2012}. In this approach, we write the magnetostatic field in a (cubic) cell, $i$ ($\bm{H}_{d,i}$), as:
\begin{equation}
\bm{H}_{d,i} = -M_s \displaystyle\sum_{j} \bm{N}(\bm{r}_i - \bm{r}_j) \cdot \bm{m}_j
\label{eq:rsc}
\end{equation}
where $\bm{N}$ is the 3$\times$3 symmetric demagnetizing tensor. The sum runs over all cells at positions $\bm{r}_{i,j}$. The demagnetizing tensor is given by:
\begin{equation}
\bm{N}(\bm{r}_i-\bm{r}_j) = \frac{1}{4\pi} \oint_{S_i} \oint_{S_j} \frac{d \bm{S}_i \cdot d \bm{S}_j^{'}}{\lvert \bm{r}-\bm{r}^{'}\rvert},
\label{eq:intmat}
\end{equation}
$S_i$ ($S_j$) are the surface of cell $i$ ($j$) respectively, $\bm{r}$ and $\bm{r}^{'}$ are the points on the surface $i$ and $j$. This sum requires a summation from all cells and requires integration over each of the surfaces $i$ and $j$, making it extremely computationally expensive. If one were to perform the integration~\ref{eq:intmat} numerically for each surface of each cell the calculation is extremely time-consuming and converges very slowly with the number of mesh points on each surface. To that end we have employed the method of Newell~\cite{Newell1993}, whereby the surface integration of the cubes is calculated analytically as in the OOMMF code~\cite{Oommf}. Some further details can be found in appendix~\ref{sec:app1}.

\section{Ferromagnetic Resonance in Thin Films of FePt}
\label{sec:FMRThinFilm}

In this section we present calculations of thin films of FePt. We begin by looking at the effect of temperature on the damping of 2~nm thin films using the stochastic form of the LLB equation with demagnetizing fields (equation~\ref{eq:rsc}). We compare this to the results for the single spin analytic results. The thin films show a large increase in the predicted damping over the analytic results due to the inclusion of the demagnetizing term as we approach the Curie temperature. The thickness dependence of the films is also calculated using the multi-spin model, showing that at low temperatures there is little variation in damping with film thickness though, at temperatures approaching the Curie point there is a large reduction with increasing thickness.

\begin{figure}[htbc!]
\begin{center}
\includegraphics[width=\columnwidth]{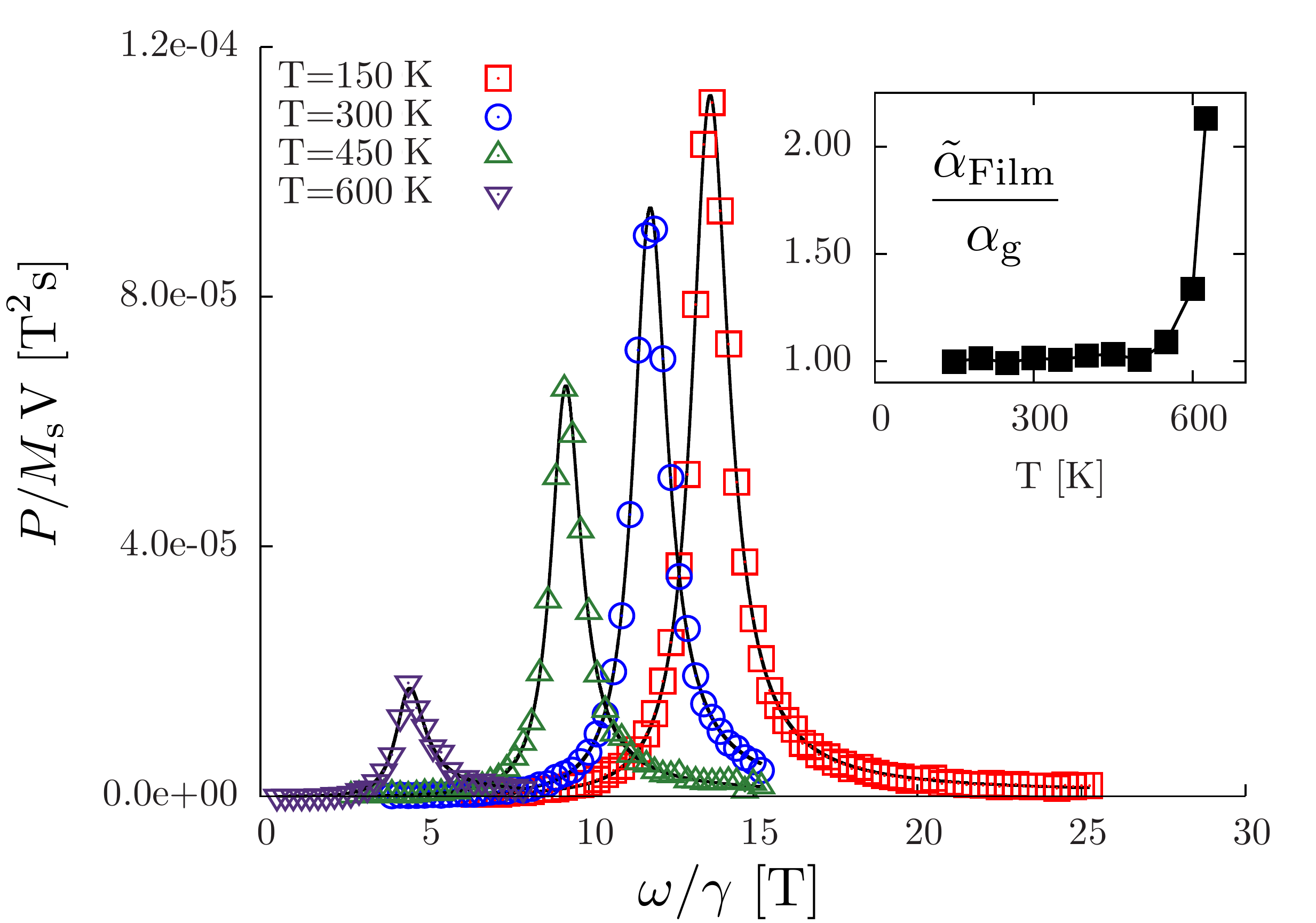}
\caption{(Color online) Ferromagnetic resonance curves in thin films of FePt for a range of temperatures below the Curie temperature. The points here are simulated data and the lines are the fits to equation~\ref{eq:fit}. The inset shows the ratio of the damping as measured in our 2D film to the damping calculated analytically for a single macrospin. For low temperatures the two are equivalent, however, at higher temperatures there is an enhanced damping in the thin films due to the effect of the magnetostatic field.}
\label{fig:film}
\end{center}
\end{figure}

By systematically varying the anisotropy we have shown that the this increase in damping occurs when the demagnetizing field dominates over the anisotropy term. Finally, this modification in the damping is shown to affect the switching times as we transition from one regime to another.

The $x$ and $y$ dimensions of the thin films in this section are 0.4$\mu$m$\times$0.4$\mu$m. The $z$-dimension is initially one cell (2~nm) thick, i.e. a 2D film. Our cell discretization is 2~nm$\times$2~nm$\times$2~nm, below the domain wall width. We apply the fields in the same orientation as discussed above. The resonance curves are shown on figure~\ref{fig:film} for a range of temperatures for the 2D (2~nm thick) film. From each FMR curve we have used a fitting procedure, as in figure~\ref{f:Bz_0T}, to calculate the damping in the 2D structures (solid lines). The inset of figure~\ref{fig:film} shows then the ratio of the damping that we calculate for the 2D structures to the analytically derived damping for single domain particles in section~\ref{sec:pw}. In the low temperature limit this ratio is consistent with the analytic solution for a single macrospin (i.e. it is 1), for high temperature however the damping is increased as the demagnetizing effects start to dominate over the anisotropy.

\begin{figure}[htbc!]
\begin{center}
\includegraphics[width=\columnwidth]{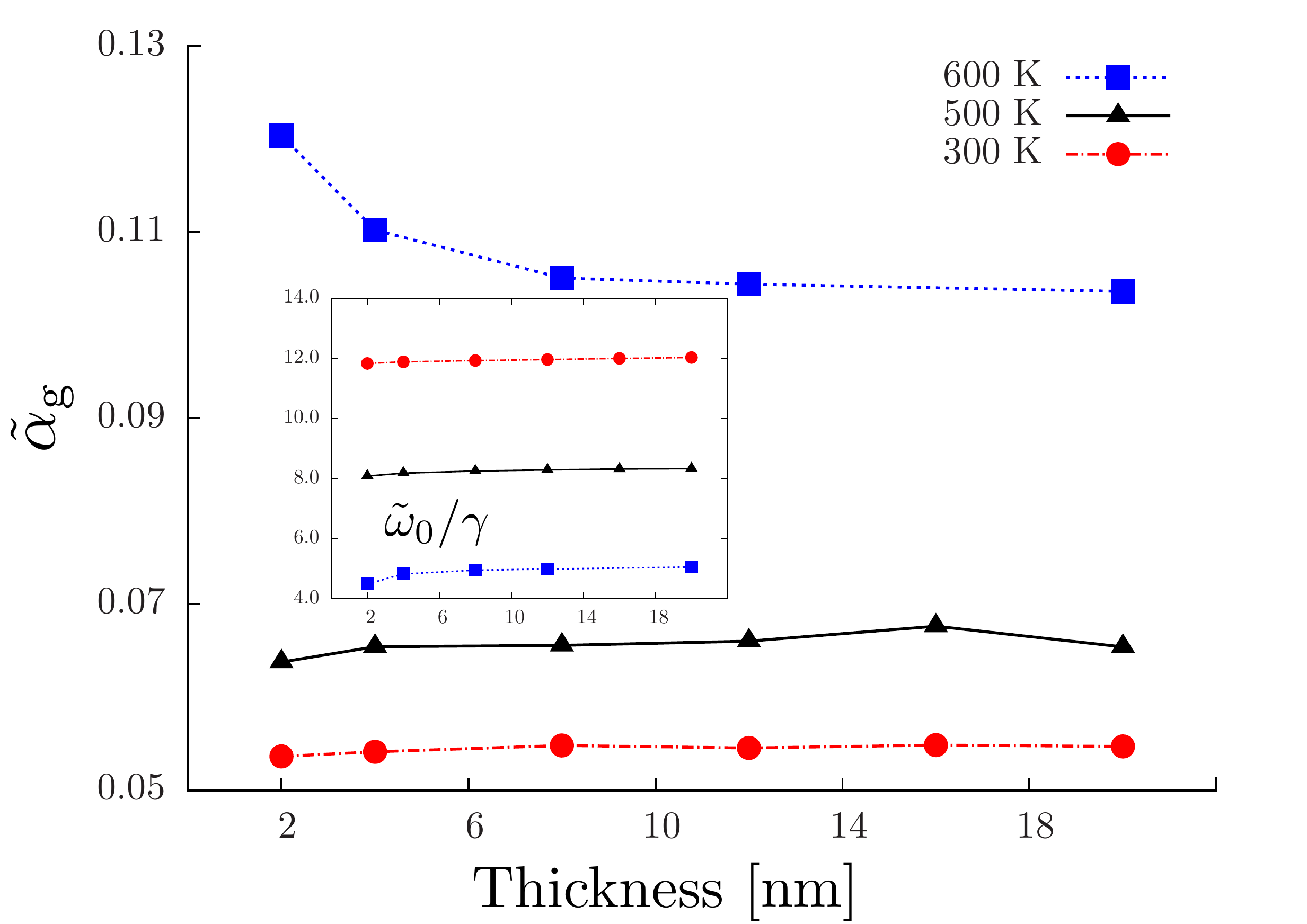}
\caption{(Color online) Damping as a function of film thickness for a range of temperature. In the low temperature regime there is a slight increasing damping as a function of thickness. As the Curie temperature is approached there is a large decrease in the damping with film thickness. The inset shows the variation of the resonance frequency with thickness. The resonance frequency shows an overall increase over all temperatures due to the decrease in the effective magnetostatic field.}
\label{fig:tdalpha}
\end{center}
\end{figure}

Next, we consider the effects of film thickness on the damping and resonance frequency. We increase the thickness of the film from 2~nm to 20~nm (1 cell to 10 cells) and calculate the ferromagnetic resonance curve for each thickness (a maximum of 400,000 cells for around 100~ns). The resulting FMR curves were again analysed to extract the damping and resonance frequencies.

Figure~\ref{fig:tdalpha} shows the variation of the damping and resonance frequency as a function of the thickness of the thin film. The largest variation in the damping is shown close to the Curie temperature (blue square, dotted line). For $T$=500~K, there is a small increase in the damping with film thickness when going from 2~nm to 4~nm. After 4~nm the curve shows little variation, consistent with the T=300~K (red circles, dot-dash line) line. The variation in the damping, with film thickness, close to the Curie point will have a large effect on the magnetization dynamics in heat assisted magnetic recording (for which FePt is a promising candidate), that operates at elevated temperatures to allow for the reduction in the anisotropy at write head field of around 1-2~T. This reduction in damping for thick layers of FePt would lead to longer switching times (as we show below), limiting the write times.

In Ref.~\onlinecite{Liu2011}, Liu \emph{et al.} showed that the damping in a magnetic tunnel junction consisting of a FeCoB free layer increased with decreasing thickness. The mechanism was said to be caused by spin pumping and nonlocal background effects. Our results, whilst are not calculated for FeCoB, show that it is not required to invoke a mechanism via spin pumping but can arise due to an interplay between the anisotropy and the demagnetizing fields.

\begin{figure}[htbc!]
\begin{center}
\includegraphics[width=\columnwidth]{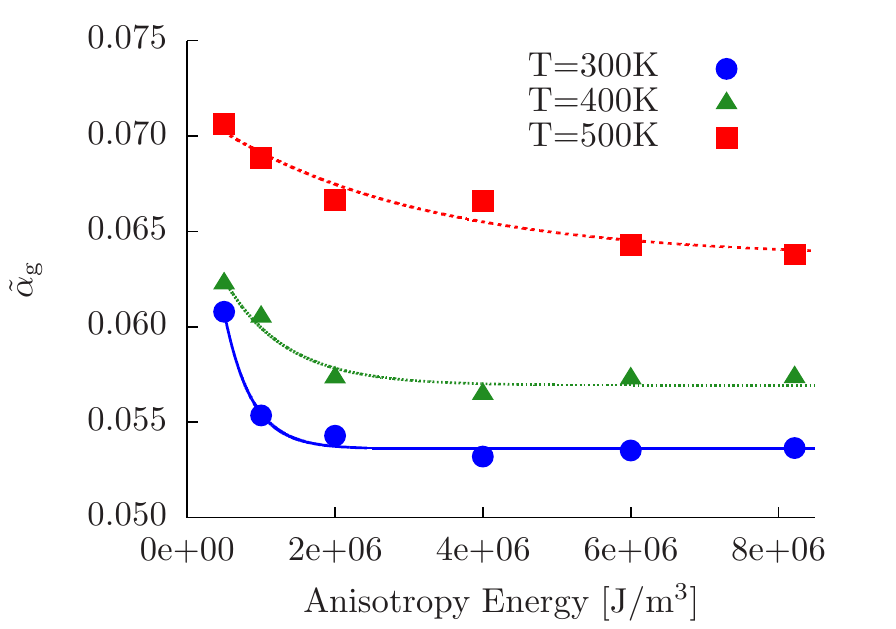}
\caption{(Color online) Dependence on the damping in FePt for a range of anisotropy constants for three values of temperature (300~K blue circle points, 400~K green triangle points and 500~K red square points). In the lower anisotropy range the damping increases, consistent the results of figure~\ref{fig:tdalpha}. The lines are fits to exponential decays to give a guide to the eye.} 
\label{fig:anis}
\end{center}
\end{figure}

As well as looking at the effect of the film thickness on the damping parameter we have also performed a systematic variation of the anisotropy constant. In FePt, the anisotropy can be modified, for example, by inducing lattice distortion or chemical disorder~\cite{Aas2011}. For the 2~nm thick films we have calculated the FMR spectra at three different temperatures (300~K, 400~K and 500~K) for a range of anisotropy values below the bulk value (vertical dashed grey line in fig.~\ref{fig:tdalpha}). From these calculations we have measured the effective damping parameters using the method described above. The overall trend shows a decrease in the measured damping, the result of which is shown in figure~\ref{fig:anis}.

The overall trend in figure~\ref{fig:anis} shows a decrease in the damping when the anisotropy becomes dominant over the demagnetizing field, consistent with the results of figure~\ref{fig:tdalpha}.

\begin{figure}[htbc!]
\begin{center}
\includegraphics[width=\columnwidth]{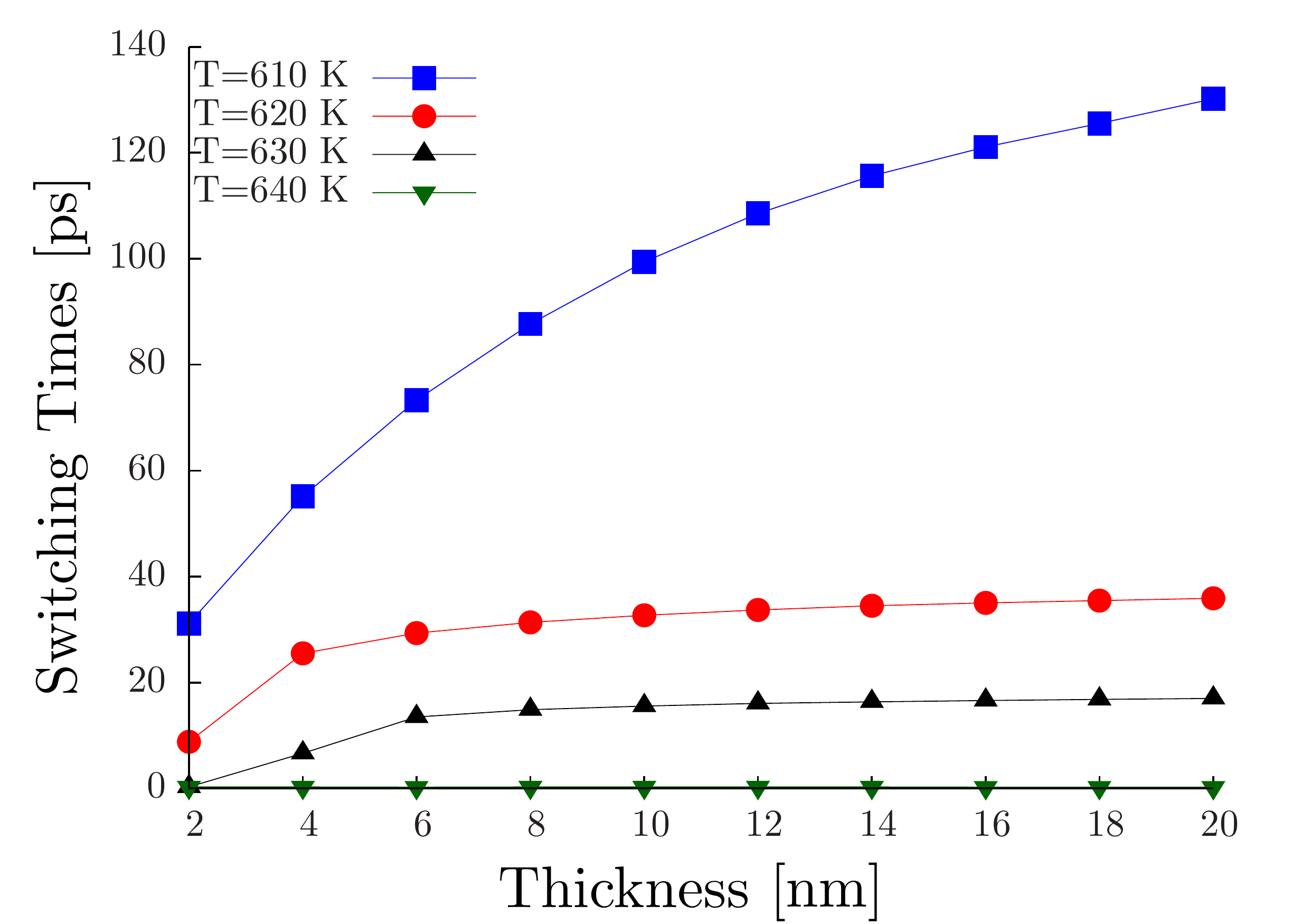}
\caption{(Color online) Switching times for thin films of FePt of differing thicknesses for a range of temperatures. Consistent with the result of figure~\ref{fig:tdalpha} the reduction in the damping with increasing film thickness leads to an increase in the switching time. A Heaviside step function of 2~T was applied to the field to reverse the magnetization after equilibration and the runs were averaged over 25 realizations of the random number seed. With the inclusion of the stochastic term there is a reduced $T_{\mathrm{C}}$ so the T=640~K line is already above the transition temperature.}
\label{fig:switch_times}
\end{center}
\end{figure}

Figure~\ref{fig:switch_times} shows the calculated switching times for four temperatures (610~K, 620~K, 630~K and 640~K) as a function of the thickness of the film. To calculate the switching times we equilibrated the system at the temperature shown in the figure, we then applied a field with a step function to 2~T to reverse the magnetization in the $z$ direction. The switching times were then averaged over 25 runs per point. The errors in the switching times are quite small so 25 runs seems to be a sufficient number to take a good average.

The thickness dependence of the switching times shown in figure~\ref{fig:switch_times} are consistent with the calculations of the damping presented in figure~\ref{fig:tdalpha}. As the thickness is increased there is an observed decrease in the damping which leads to the reduced switching times seen in figure~\ref{fig:switch_times}. It should be pointed out that the field that we apply is not sufficient to switch the magnetization below around 610~K, consistent with Ref.~\onlinecite{Kazantseva2009}. The large reduction in the switching time seen for the $T$=640~K case is due to the fact that with the inclusion of the stochastic term there is a slight reduction in the Curie temperature as shown in Ref.~\onlinecite{Evans2012}.

\section{CONCLUSION}
\label{s:conc}

We have derived, using the LLB formalism an analytic solution to the power frequency spectrum for nano metre-sized, single domain ferromagnets during ferromagnetic resonance. Using the technologically relevant FePt, this analytic solution agrees well with numerical simulations of both single spin and exchange-coupled multi-spin calculations including the stochastic thermal term.

Analysis of the resulting FMR expressions for a single macrospin show that the analytically derived damping is consistent with those of extended thin films up to quite high temperatures. At temperatures close to $T_{\mathrm{C}}$, the anisotropy decreases quickly and the demagnetizing effects play a more dominant role for the tin films. This means that our analytic expressions for thin films of magnetically soft materials would not hold, however, the analysis is still valid for single macrospins (or small structures) of soft materials.

We have extended the calculations to include the thermal stochastic term and demagnetizing effects to explore the effect this plays on the temperature dependent ferromagnetic resonance curves. By calculating FMR spectra as a function of film thickness, we have shown that there is an increased damping for thinner films due to the interplay between the demagnetizing fields and the anisotropy. For the thinner films there is more of a tendency for the films to want to lie in-plane due to the demagnetizing field. For highly anisotropic materials (shown here for FePt) this effect is more dominant at elevated temperatures. We have verified that this increase in damping can be explained by change in the dominance of the demagnetizing energy by varying the anisotropy constant for the thin films. As the anisotropy constant is decreased the damping increases, consistent with the results of varying the film thickness.

Finally, we have shown that this reduction in the damping has an effect on the switching times. This conclusion could have important consequences for heat assisted magnetic recording, which operates at elevated temperatures, and require sufficiently thick grains to have sufficient material for good read back of the magnetic signal.

\section*{Acknowledgements}

This work was supported by the European Commission under contract No. 281043, \emph{FemtoSpin}. The financial support of the Advanced Storage Technology Consortium is gratefully acknowledged.

\appendix

\section{Details of Analytic Derivation of P($\omega$)}
\label{sec:app2}

This section gives some more detail regarding the derivation of the key equations discussed in section~\ref{sec:pw}. The linearised equations of motion for the LLB equation~\ref{e:llb} are written:

\begin{eqnarray}
\dot{m}_x   \approx  &-&\gamma (m_y H_{\rm eff}^z - m H_{\rm eff}^y) \nonumber \\
                                    &  +& \frac{\gamma (\alpha _{||}-\alpha_{\perp})}{m}(m_x H_{\rm eff}^z)  \nonumber \\
                                   &+ &\frac{\gamma \alpha_{\perp}}{m}(m H_{\rm eff}^x ) \nonumber \\
\dot{m}_y  \approx & - &\gamma (m H_{\rm eff}^x - m_x H_{\rm eff}^z) \nonumber \\
          &+& \frac{\gamma (\alpha _{||}-\alpha_{\perp})}{m}(m_y H_{\rm eff}^z) \nonumber \\
          &+& \frac{\gamma \alpha_{\perp}}{m}(m H_{\rm eff}^y ) \nonumber \\
\dot{m}_z    &=& 0,
\label{e:linLLB}
\end{eqnarray}
with the linearised effective fields then written as:
\begin{eqnarray}
  \bm{H}_{\mathrm{eff}}^{x,y} &=& B_{x,y} -\frac{m_{x,y}}{\tilde{\chi}_{\perp}}
  +     \frac{1}{2\tilde{\chi}_{\Vert }}\left(1-\frac{m^{2} }{m_{\rm e}^{2}}\right) m_{x,y}  \nonumber \\
  \bm{H}_{\mathrm{eff}}^z &=& B_z  +
     \frac{1}{2\tilde{\chi}_{\Vert }}\left(
     1-\frac{m^{2} }{m_{\rm e}^{2}}\right) m .
\label{e:linHeff}
\end{eqnarray}
In equilibrium the $z$-component of the effective field vanishes, $ \bm{H}_{\mathrm{eff}}^z = 0$.  
Using the linearised form of $m$, $m = m_{\rm e} (1 + \Delta m)$ as well as  $m^2 = m_{\rm e}^2 (1 + 2 \Delta m)$, 
we arrive at an expression for the $z$-component of the applied magnetic field
\begin{equation}
B_z  - \frac{m_{\rm e} \Delta m +  m_{\rm e} \Delta m^2}{\tilde{\chi}_{\Vert }}   = 0 . \nonumber \\
\label{e:Bz}
\end{equation}
Using the linearised form of this equation, 
$B_z  - \frac{m_{\rm e} \Delta m }{\tilde{\chi}_{\Vert }}   = 0 $
as well as the approximation $\Delta m = (m-m_{\rm e})/m_{\rm e}$ we have an approximation for
$m$ during FMR that is both field and temperature dependent:
\begin{equation}
 m(T,B_z)  =  \tilde{\chi}_{\Vert}(T) B_z + m_{\rm e}(T).
 \label{e:approxm}
\end{equation}

As is discussed in the main text the approximation~\ref{e:approxm} leads to errors in the analytic treatment if the resonance curve is calculated in an applied field. This is due to the fact that the susceptibility diverges as we approach the Curie temperature. This does not occur in the numerical simulations and is only a problem in the analytic calculations due to the above approximation~\ref{e:approxm}.

In order to calculate the resonance frequency ($\omega_0$) as well as the transverse relaxation time ($\tau$) for the power spectrum $P(\omega)$ one has to solve the 
linearised LLB equation (see Eq.~\ref{e:linLLB}). Using the notation $\tilde{m} = m_x + {\rm i}m_y$ and 
$\tilde{H}_{\rm eff} =  H_{\rm eff}^x  + {\rm i} H_{\rm eff}^y$ leads to the differential equation,
\begin{equation}
 \frac{\dot{\tilde{m}}}{\gamma} = \tilde{m} \left({\rm i} + \frac{\alpha _{||}-\alpha_{\perp}}{m}\right) {H}^z_{\rm eff}
                          + m \left(\frac{\alpha_{\perp}}{m} - {\rm i}\right)\tilde{H}_{\rm eff} . \nonumber
\label{e:DGLm}                          
\end{equation}
As can be easily seen from Eq.~\ref{e:linHeff}, $\tilde{H}_{\rm eff}$ is also $\tilde{m}$-dependent. Writing the effective field as, $\tilde{H}_{\rm eff}= \tilde{B} + A \tilde{m}$,  with 
 $A = -\frac{1}{\tilde{\chi}_{\perp}} + \frac{1}{2\tilde{\chi}_{\Vert }}( 1-\frac{m^{2}}{m_{\rm e}^2})$ and   $\tilde{B} = B_x + {\rm i} B_y$ we arrive at an inhomogeneous differential equation:
\begin{eqnarray}
\frac{\dot{\tilde{m}}}{\gamma} &=&  \tilde{m} \left(({\rm i} + \frac{\alpha _{||}-\alpha_{\perp}}{m}) H_{\rm eff}^{z} \right) \nonumber \\
                             &+& \tilde{m} \left( m (\frac{\alpha_{\perp}}{m} - {\rm i})A \right) \nonumber \\
&+& m( \frac{\alpha_{\perp}}{m} - {\rm i}) \tilde{B}.
\label{e:DGLm1}                          
\end{eqnarray} 
In the first step, we solve the homogeneous part of the differential equation \ref{e:DGLm1}, using the Ansatz $\tilde{m}_{\rm{hom}}(t) = \exp{(\omega t)}$ whose solution leads to the expressions for $\omega_0$ and $\tau$:
\begin{eqnarray}
 \omega_0  &=& \gamma \Big(B_z  + \frac{m}{\tilde{\chi}_{\perp}}\Big) 
 \label{e:ares1}
 \end{eqnarray}
 \begin{eqnarray}
\tau & =& \frac{m}{\lambda \big( (1 - \frac{T}{3 \TC})\omega_0 - \frac{2}{3} \gamma \frac{T}{\TC} {H}_{\mathrm{eff}}^z \big)}.
 \label{e:ares2}
\end{eqnarray}
In the next step, we solve the inhomogeneous differential equation \ref{e:DGLm1} under  the assumption that the applied magnetic field ${\bf B}$ has the form ${\bf B} = (B_0 \exp({\rm i}\omega t) , 0 , B_z)$, where $B_0 \ll B_z$.  
These lead to the following simplification of the right hand side of Eq.~\ref{e:DGLm1},
\begin{equation}
m (\frac{\alpha_{\perp}}{m} - {\rm i}) \tilde{B}  = m (\frac{\alpha_{\perp}}{m} - {\rm i})  B_0 \exp({\rm i}\omega t).
\end{equation}
Using the Ansatz $\tilde{m}(t) = u(t) \tilde{m}_{\rm hom}(t)$ where $u(t)$ is given by
\begin{eqnarray}
u(t) &=& \int_{t_0}^t \frac{m (\frac{\alpha_{\perp}}{m} - {\rm i}) B_0 \exp({\rm i}\omega t)}{\exp(-\frac{t}{\tau}) \exp({\rm i}\omega_0 t)}{\rm d}t, \nonumber
\end{eqnarray}
and assuming $t_0 = 0$ and $t \rightarrow \infty$ Eq.~\ref{e:DGLm1} has the solution
\begin{equation}
\tilde{m}(t) = \frac{(-{\rm  i} + \frac{\alpha_{\perp}}{m})\gamma m  B_0 (\frac{1}{\tau} - \rm{i}(\omega - \omega_0))}{\frac{1}{\tau ^2}+ (\omega - \omega_0)^2} \exp({\rm i} \omega t) . \nonumber 
\end{equation}
From this general solution, $m_x$ can easily be derived
\begin{eqnarray} 
m_x &=&\frac{\gamma m B_0}{\frac{1}{\tau^2} + (\omega - \omega_0)^2} \Bigg( \left(\frac{\alpha_{\perp}}{\tau m}-(\omega-\omega_0)\right)\cos(\omega t) \nonumber \\
&+& \left(\frac{1}{\tau}+\frac{\alpha_{\perp}}{m}(\omega - \omega_0)\right) \sin(\omega t)\Bigg), \nonumber \\
\label{e:mx}
\end{eqnarray}
and substituted into the definition for the power spectrum $P(\omega)$ (see Eq.~\ref{e:defP}). This leads to the analytic solution for the power spectrum $P(\omega)$:
\begin{eqnarray}
P(\omega)  &=& \frac{\mu_{\rm s} \omega^2}{4}  \frac{\gamma \alpha_{\perp} B_0^2}{\frac{1}{\tau^2} + (\omega - \omega_0)^2}.
\label{e:LLBP_redAPP}
\end{eqnarray}

As we can see from equation~\ref{e:LLBP_redAPP}, the analytic solution for the absorbed power depends on the magnetization, which in-turn depends on the longitudinal susceptibility. As mentioned above, we approximate the magnetization in the presence of an applied field (equation~\ref{e:approxm}) in terms of the zero field susceptibility. Therefore, equation~\ref{e:approxm} is only strictly correct in the zero field limit. Away from the critical temperature the zero field susceptibility is small, therefore in this limit the approximation holds. As we approach the critical temperature the susceptibility diverges as we approach the phase transition. This means that our analytic expression shows a deviation from the numerically calculated result.

\begin{figure}[htbc!]
\begin{center}
\includegraphics[angle=0,width=\columnwidth]{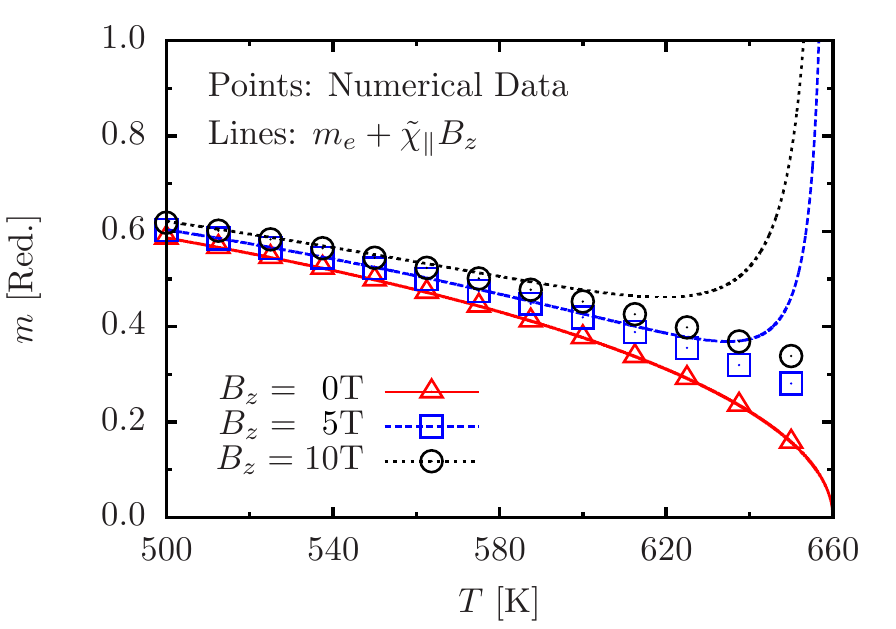}
\caption{(Color online) Equilibrium magnetization vs temperature in different applied fields. The (red) solid line represents the zero-field equilibrium magnetization, $m_{\rm e}$, gained from atomistic FePt simulations, a fit to which defines the input function, $m_e(T)$~\cite{Kazantseva2008}. The dashed (blue) and dotted (black) lines represent expression~\ref{e:approxm} for different applied fields. The symbols represents the equilibrium magnetization in the presence of an applied field, $B_z = 0,5,10$T, from the numerical simulations of a single macrospin without demagnetizing, stochastic or exchange fields.}
\label{f:M_T}
\end{center}
\end{figure}

A plot of the magnetization as a function of temperature using equation~\ref{e:approxm} and data from numerical simulations can be seen in Figure~\ref{f:M_T}. For small values of the applied field this error reduces as the susceptibility is defined for small changes in the applied field.

Figure~\ref{f:M_T} shows the equilibrium magnetization (red solid curve), initially calculated from atomistic spin dynamics simulations~\cite{Kazantseva2008}, which is used as an input to the numeric simulation. As well as the equilibrium magnetization, the figure also shows the magnetization as a function of temperature in 5 and 10T applied fields, which is of course not zero at the (zero field) Curie temperature. The dashed and dotted line is the analytic solution to the magnetization (also in 5 and 10T fields), diverging across the Curie temperature. As we can see, the magnetization in an applied field from the analytic expression shows a diverging behaviour as we approach the Curie temperature because of the diverging susceptibility, whereas the numerical simulations (points) show no such divergence.

\section{Magnetostatic Fields}
\label{sec:app1}

For efficient calculation of the magnetostatic fields we write the convolution~\ref{eq:rsc} as:
\begin{equation}
H_{d,i}^{\eta} = \displaystyle\sum_{\theta,j}W_{ij}^{\eta \theta} m_j^{\theta}
\label{eq:rsc2}
\end{equation}
where the Greek symbols $\eta$, $\theta$ again denote Cartesian components $x$,$y$,$z$ and Latin symbols $i$, $j$ denote the lattice sites. $W_{ij}^{\eta \theta}$ are interaction matrices which only depend on the structure of the material (cubic in this work). Since we are considering a translationally invariant lattice one can apply the discrete convolution theorem and calculate the fields in Fourier space:
\begin{equation}
H_{d,k}^{\eta}=\displaystyle\sum_{\theta} W_{k}^{\eta \theta} m_k^{\theta}.
\label{eq:fsc}
\end{equation}
It should be pointed out here that we have absorbed the prefactor, $M_s$ into the interaction matrix, $W_{ij}^{\eta \theta}$. Furthermore to write the fields in terms of units of Tesla to be consistent with the form of the fields above, we have multiplied equation~\ref{eq:rsc} by $\mu_0$. The Fourier transform of the interaction matrix only has to be performed once and thus stored in memory. 

There are a number of conditions that must be met in order to utilize the convolution theorem. In terms of signal processing theory the interaction matrix is seen as the response function and the magnetization data is the signal. We should not that there are two conditions that must be satisfied to utilize the convolution theorem. The first is that the signal (spin system) must be periodic in space. The second is that the range of the response function should be the same as the signal~\cite{Berkov1998}. The magnetic system is usually not periodic and the demagnetizing effects are long ranging and cannot be cut-off at a reasonable distance due to the slow decay~\cite{Berkov1998}. To solve this we simulate a finite system, therefore to meet the above requirements it is required that we zero pad the magnetization configurations by doubling the size of each dimension and adding zero's in the areas where there are no macrospins.

At each update of the demagnetizing field (every 10fs) the Fourier transform of the magnetization arrays is performed and the resulting Fourier components convoluted with that of the interaction matrix. The resulting product is back transformed via an inverse Fourier transform to give the demagnetizing fields in real space.

\end{document}